\def\tr{{\text{tr}}\,}
\def\Tr{{\text{Tr}}\,}

\def\be{\begin{equation}}
\def\ee{\end{equation}}
\def\bea{\begin{eqnarray}}
\def\eea{\end{eqnarray}}
\def\bse{\begin{subequations}}
\def\ese{\end{subequations}}

\documentclass[prb,twocolumn,showpacs,amsmath,amssymb,eqsecnum]{revtex4}
\usepackage{graphicx}
\usepackage{dcolumn}
\usepackage{bm}
\begin{document}
\title{Split transition in ferromagnetic superconductors
}
\author{D.Belitz}
\affiliation{Department of Physics and Materials Science Institute\\
         University of Oregon,
         Eugene, OR 97403}
\author{T.R.Kirkpatrick}
\affiliation{Institute for Physical Science and Technology, and Department of
         Physics\\
         University of Maryland,
         College Park, MD 20742}
\date{\today}

\begin{abstract}
The split superconducting transition of up-spin and down-spin electrons on the
background of ferromagnetism is studied within the framework of a recent model
that describes the coexistence of ferromagnetism and superconductivity induced
by magnetic fluctuations. It is shown that one generically expects the two
transitions to be close to one another. This conclusion is discussed in
relation to experimental results on URhGe. It is also shown that the magnetic
Goldstone modes acquire an interesting structure in the superconducting phase,
which can be used as an experimental tool to probe the origin of the
superconductivity.
\end{abstract}

\maketitle


\section{Introduction}
\label{sec:I}

Recently, the coexistence of ferromagnetism and superconductivity has been
observed in a number of materials, including,
UGe$_2$,\cite{Saxena_et_al_2000,Huxley_et_al_2001} URhGe,\cite{Aoki_et_al_2001}
and ZrZn$_2$.\cite{Pfleiderer_et_al_2001} The experiments so far have
ascertained the presence of bulk superconductivity, and various thermodynamic
and transport properties have been measured, but little is known yet about the
detailed nature of the superconducting state. An obvious possibility is
spin-triplet pairing induced by ferromagnetic fluctuations, but other
mechanisms have also been proposed.\cite{pairing_mechanisms_reference} One
interesting aspects of the experiments, which may provide a clue about the
origin and the nature of the pairing, is that the superconductivity is observed
only on the ferromagnetic side of the magnetic phase boundary. This is in sharp
contrast to early theories of superconductivity induced by ferromagnetic
fluctuations, which predicted that this type of superconductivity would be
equally strong on the paramagnetic side of the magnetic phase
boundary.\cite{Fay_Appel_1980} However, in a recent paper,\cite{us_p-wave} to
be referred to as I, it has been shown that the presence of magnons in the
ferromagnetic phase can lead to a drastic enhancement of the superconducting
$T_{\text{c}}$ compared to the paramagnetic phase. Clearly, more properties of
the superconducting state must be studied in order to differentiate between
different possible pairing mechanisms. In this context, an interesting result
is the recent measurement of the specific heat in URhGe.\cite{Aoki_et_al_2001}
Taken at face value, this experiment shows a single transition into a
superconducting state, and a low temperature specific heat that is linear in
the temperature, which suggests that a fraction of the electrons remain in a
Fermi-liquid state at low temperatures. Similar behavior has been observed in
UGe$_2$.\cite{Tateiwa_et_al_2001}

To date all of the theoretical work has assumed only the simplest type of
superconducting order for a given pairing mechanism, with the main goal being
to determine the phase boundary for superconductivity. For example, in I the
present authors assumed an ordering of spin-triplet Cooper pairs with spins
oriented in the direction of the magnetization. Let us denote the gap function
for this ordering by $\Delta_{\uparrow}$. Previous work on Helium-3 in a
magnetic field\cite{Vollhardt_Woelfle_1990} suggests that at some point, the
Cooper pairs with spins oriented opposite to the magnetization, characterized
by a gap function $\Delta_{\downarrow}$, will form as well. More generally, the
complete phase diagram for these systems will likely be complicated and involve
ferromagnetism coexisting with several types of superconducting
order.\cite{flux_lattice_footnote}

Our goal in this paper is three-fold. First, we develop a formal theory that
enables us to consistently describe ferromagnetism coexisting with two types of
superconducting order. The superconductivity in our theory is caused by
ferromagnetic fluctuations. Our goal is to derive an equation of state,
analogous to the strong-coupling or Eliashberg equations of conventional
superconductivity, that describes both up-spin and down-spin superconducting
order, as well as a consistent magnetic equation of state in the presence of
superconductivity. A major complication is that, because the superconductivity
is itself caused by a fluctuation effect, a simple mean-field theory is not
sufficient, and fluctuations need to be taken into account. The fluctuations
that cause superconductivity within our theory are described by the spin
susceptibility tensor. Our second goal is therefore to develop a theory for the
spin susceptibility, both in a pure ferromagnetic phase, and in coexisting
ferromagnetic and superconducting phases. The pairing potential for
superconductivity involving only $\Delta_{\uparrow}$ and $\Delta_{\downarrow}$
is given by the longitudinal susceptibility, $\chi_{\text{L}}$. However, the
transverse susceptibilities, $\chi_{\text{T}}$, are also needed because they
enter the normal self energies, and because they couple to $\chi_{\text{L}}$
via mode-mode-coupling effects. In fact, as shown in I, within the framework of
our theory it is this mode-mode-coupling mechanism, which exists only in a
magnetically ordered state, that causes the superconductivity to be observable
only in the ferromagnetic phase. We will see that the magnons, which are
described by $\chi_{\text{T}}$, have an interesting dispersion relation in the
superconducting state and, under certain conditions, can become effectively
massive. A striking feature of the mass is that it is proportional to the
inverse square of the magnetization, and thus singular for small magnetization.
We give estimates showing that this result should be observable with neutron
scattering.

\begin{figure}[t]
\vskip 10mm
\includegraphics[width=6cm]{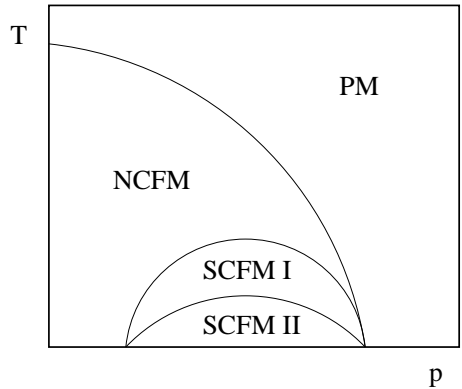}
\caption{\label{fig:1} Schematic phase diagram in the temperature-pressure
  plane showing a paramagnetic phase (PM), a normal conducting ferromagnetic one
  (NCFM), a superconducting ferromagnetic phase with up-spin pairing only (SCFM
  I), and a superconducting ferromagnetic phase with both up-spin and down-spin
  pairing (SCFM II). The two superconducting phases have been drawn to end in the
  same point for simplicity, but this will not necessarily be the case.}
\end{figure}

Our third goal is to discuss the phase diagram for superconductivity and
magnetism, based on the above results. In a theory that allows for pairing of
both up-spin and down-spin electrons, one expects a phase diagram that is
qualitatively shown in Fig.\ \ref{fig:1}. With decreasing temperature, one
oberves a transition from a paramagnetic (PM) phase to a normal conducting
ferromagnetic (NCFM) one. With decreasing temperature, the system enters a
ferromagnetic superconducting state where only up-spin electrons are paired
(SCFM I). Finally, with further decreasing temperature, down-spin electrons are
paired as well (SCFM II). In general the coupled magnetic and superconducting
equations of state are very complicated to solve. For the points we want to
make, however, a complete solution is not needed. Our aim here is to compute,
for our proposed pairing mechanism, the relative magnitudes of the transition
temperatures $T_{\text{c}\uparrow}$ and $T_{\text{c}\downarrow}$ for the phase
boundaries between the NCFM phase and the SCFM I phase, and between the SCFM I
phase and the SCFM II phase, respectively. Generically we find that these
transition temperatures are close to one another. Experimentally this suggests
that, for example, any specific heat measurement should find two closely spaced
transition signatures. This result is in conflict with the naive interpretation
of the specific heat measurement in URhGe noted above. A possible alternative
interpretation of the data consistent with our theory will be discussed below.

The plan of this paper is as follows. In Section \ref{sec:II} we develop a
formalism that allows for a consistent description of all components of
spin-triplet superconductivity, induced by ferromagnetic fluctuations, in the
presence of long-range ferromagnetic order. In Section \ref{sec:III} we
calculate the magnetic susceptibility in the ferromagnetic state in the
presence of a non-unitary superconducting order parameter. In Section
\ref{sec:IV} we solve the strong-coupling equations for superconductivity and
determine the phase diagram containing phases with pure ferromagnetic order,
ferromagnetic plus spin-up superconducting order, and ferromagnetic plus both
spin-up and spin-down superconducting order, respectively. We discuss our
results and their experimental implications in Section \ref{sec:V}. In Appendix
\ref{app:A} we augment our microscopic approach with a more general
Landau-Ginzburg-Wilson theory and discuss the soft-mode structure of the
magnetic superconducting phase. In Appendix \ref{app:B} we relate the physical
spin susceptibility to the magnetization fluctuations that occur most naturally
in the theory. Some of our results have been reported before in Ref.\
\onlinecite{Kirkpatrick_Belitz_2004}.

\vskip 0cm

\section{A Field-Theoretic Approach to Superconductivity and Magnetism}
\label{sec:II}

\subsection{The model}
\label{subsec:II.A}

Our starting point is the same model of interacting electrons as in I. That is,
we consider free electrons with a static, point-like spin-triplet interaction
with amplitude $\Gamma_{\text t}$. For simplicity we ignore the spin-singlet
interaction. We do {\em not} include an explicit Cooper channel interaction;
the pairing interaction will be generated by magnetic fluctuations. We then
decouple the spin-triplet interaction by means of a Hubbard-Stratonovich
transformation. Bilinear products of fermionic fields are constrained to a
bosonic field ${\cal G}$ by means of a Lagrange multiplier field $\Lambda$ with
a transposed $\Lambda^{\text{T}}$, and we integrate out the fermions. The
resulting action is given by Eq.\ (2.12a) of I, with $\Gamma_{\text s} = 0$,
\bea
{\cal A}[{\bm M},{\cal G},\Lambda] &=& \frac{1}{2}\,\Tr\ln ({\tilde G}_{0}^{-1}
   + \sqrt{2\Gamma_{\text t}}
   {\bm\gamma}\cdot{\bm M} - \Lambda^{\text T})
\nonumber\\
&&\hskip -15pt + \Tr(\Lambda\,{\cal G}) - \int dx\ {\bm M}(x)\cdot {\bm M}(x)\
.
\label{eq:2.1}
\eea
Here $\tilde{G}_0^{-1}$ is the bare inverse Green operator,
\be
{\tilde G}_{0}^{-1} = -\partial_{\tau}
                      + \gamma_0\left({{\bm\nabla}^2}/2m_{\rm e}
                      + \mu\right)\quad,
\label{eq:2.2}
\ee
where $\mu$ is the chemical potential and $m_{\text{e}}$ is the electron mass.
We denote the real-space and imaginary-time coordinates by $\bm{x}$ and $\tau$,
respectively, combine these into a four-vector $x = (\bm{x},\tau)$, and use the
notation $\int dx = \int_V d\bm{x} \int_0^{1/T} d\tau$, with $V$ and $T$ the
system's volume and temperature, respectively. ${\bm M}$ is the
Hubbard-Stratonovich field whose expectation value is proportional to the
magnetization $m$, and ${\bm \gamma}$ denotes three components of a four-vector
of $4\times 4$ matrices,
\be
(\gamma_0,{\bm \gamma}) = (\sigma_3\otimes\sigma_0,\sigma_3\otimes\sigma_1,
                      \sigma_0\otimes\sigma_2,\sigma_3\otimes\sigma_3)\quad,
\label{eq:2.3}
\ee
with $\sigma_{1,2,3}$ and $\sigma_0$ the Pauli matrices and the $2\times 2$
unit matrix, respectively. $\mathcal{G}(x,y)$ is a $4\times 4$ matrix field
which, in contrast to $\bm{M}$, depends on two space-time variables. Its
expectation value determines the various Green functions of the electron
system. Specifically,
\bse
\label{eqs:2.4}
\bea
\langle\mathcal{G}_{11}(x,y)\rangle &=& - \langle\mathcal{G}_{33}(x,y)\rangle
   =
   \frac{1}{2}\,\langle{\bar\psi}_{\uparrow}(x)\psi_{\uparrow}(y)\rangle_{\psi}
\nonumber\\
&\equiv&\frac{1}{2}\,G_{\uparrow}(x-y)\quad,
\label{eq:2.4a}\\
\langle\mathcal{G}_{22}(x,y)\rangle &=& - \langle\mathcal{G}_{44}(x,y)\rangle
   = \frac{1}{2}\,\langle{\bar\psi}_{\downarrow}(x)\psi_{\downarrow}(y)
                                                                 \rangle_{\psi}
\nonumber\\
     &\equiv&\frac{1}{2}\,G_{\downarrow}(x-y)\quad,
\label{eq:2.4b}
\eea
are the normal Green functions, while
\bea
\langle\mathcal{G}_{13}(x,y)\rangle &=& \frac{1}{2}\,
\langle{\bar\psi}_{\uparrow}(x){\bar\psi}_{\uparrow}(y)\rangle_{\psi}
      \equiv\frac{1}{2}\,F^{+}_{\uparrow}(x-y)\ ,\qquad
\label{eq:2.4c}\\
\langle\mathcal{G}_{24}(x,y)\rangle &=&
\frac{1}{2}\,\langle{\bar\psi}_{\downarrow}(x){\bar\psi}_{\downarrow}(y)
       \rangle_{\psi}\equiv\frac{1}{2}\,F^{+}_{\downarrow}(x-y)\ ,
\label{eq:2.4d}\\
\langle\mathcal{G}_{31}(x,y)\rangle &=&
\frac{1}{2}\,\langle\psi_{\uparrow}(x)\psi_{\uparrow}(y)\rangle_{\psi}
      \equiv\frac{1}{2}\,F_{\uparrow}(x-y)\ ,
\label{eq:2.4e}\\
\langle\mathcal{G}_{42}(x,y)\rangle &=&
\frac{1}{2}\,\langle\psi_{\downarrow}(x)\psi_{\downarrow}(y)\rangle_{\psi}
      \equiv\frac{1}{2}\,F_{\downarrow}(x-y)\ ,
\label{eq:2.4f}
\eea
\ese%
are the anomalous ones. Here $\langle\ldots\rangle$ and
$\langle\ldots\rangle_{\psi}$ denote averages with respect to the effective
action ${\cal A}$ and the underlying fermionic action, respectively. Since we
do not allow for mixed pairing of up- and down-spins, which one expects to be
strongly suppressed, these are the only nonzero Green functions. We decompose
${\bm M}$ and ${\mathcal G}$ into their expectation values and fluctuations
\bse
\label{eqs:2.5}
\bea
\bm{M}(x) = m\sqrt{\Gamma_{\text{t}}/2}\ \hat{z} + \delta\bm{M}(x)\quad,
\label{eq:2.5a}\\
\mathcal{G}(x,y) = \langle\mathcal{G}(x,y)\rangle + \delta\mathcal{G}(x,y)
                   \quad.
\label{eq:2.5b}
\eea
Here $m$ is the magnetization, which we assume to be in $z$-direction. For the
Lagrange multiplier field $\Lambda$ we write, in analogy to Eq.\
(\ref{eq:2.5b}),
\be
\Lambda(x,y) = \lambda(x-y) + \delta\Lambda(x,y)\quad.
\label{eq:2.5c}
\ee
\ese%
The matrix elements of $\lambda$ represent the normal self energies,
\bse
\label{eqs:2.6}
\bea
\lambda_{11}(x-y)&=&-\lambda_{33}(x-y) \equiv \Sigma_{\uparrow}(x-y)\quad,
\label{eq:2.6a}\\
\lambda_{22}(x-y)&=&-\lambda_{44}(x-y) \equiv \Sigma_{\downarrow}(x-y)\quad,
\label{eq:2.6b}
\eea
and the anomalous ones,
\bea
\lambda_{13}(x-y)&\equiv&\Delta^{+}_{\uparrow}(x-y)\quad,
\label{eq:2.6c}\\
\lambda_{24}(x-y)&\equiv&\Delta^{+}_{\downarrow}(x-y)\quad,
\label{eq:2.6d}\\
\lambda_{31}(x-y)&\equiv&\Delta_{\uparrow}(x-y)\quad,
\label{eq:2.6e}\\
\lambda_{42}(x-y)&\equiv&\Delta_{\downarrow}(x-y)\quad.
\label{eq:2.6f}
\eea
\ese%
Despite the formal similarities in our treatments of $\bm{M}$ and
$\mathcal{G}$ on one hand, and $\Lambda$ on the other,
it is important to note that $\lambda$ is {\em not} the expectation value
of $\Lambda$. Rather, it will be determined self-consistently by the
method explained below.

\subsection{Expansion in powers of the fluctuations}
\label{subsec:II.B}

We now expand the action in powers of the fluctuations $\delta\bm{M}$ and
$\delta\mathcal{G}$, as well as the quantity $\delta\Lambda$. To this end it is
useful to define an inverse Green operator
\be
\tilde{G}^{-1} = \tilde{G}_0^{-1} + m\Gamma_{\text t}\gamma_3
                 - \lambda^{\text T}
\label{eq:2.7}
\ee
The zeroth-order or mean-field action then reads
\be
\mathcal{A}^{(0)} = -\,\frac{V\Gamma_{\text{t}}}{2T}\,m^2
   + \Tr (\lambda \langle\mathcal{G}\rangle) + \frac{1}{2}\Tr\ln\tilde{G}^{-1}
   \quad.
\label{eq:2.8}
\ee
This mean-field action or Landau theory, and its generalization to a
Landau-Ginzburg-Wilson (LGW) theory, are interesting in their own right, and we
discuss them in Appendix \ref{app:A}. To bilinear order in the fluctuations, it
is obvious from Eq.\ (\ref{eq:2.1}) that $\delta\Lambda$ couples to both
$\delta G$ and $\delta{\bm M}$. This coupling can be eliminated by first
shifting $\delta{\bm M}$, and then $\delta\Lambda$. The diagonalized Gaussian
action then takes the form
\bea
\mathcal{A}^{(2)}&=&- \left(\delta\bm{M}\vert\chi^{-1}\vert\delta\bm{M}\right)
  - \frac{1}{2}\left(\delta\Lambda\vert\Gamma\vert\delta\Lambda\right)
\nonumber\\
&&  + \frac{1}{2}\left(\delta\mathcal{G}^{\text{T}}\vert\Gamma^{-1}\vert
        \delta\mathcal{G}^{\text{T}}\right)\quad.
\label{eq:2.9}
\eea
Here we employ an obvious scalar product notation that implies summation
and integration over all discrete and continuous indices, respectively,
of the various fields. $\chi^{-1}$ is an inverse Gaussian magnetic
susceptibility defined by
\bse
\label{eqs:2.10}
\be
\chi^{-1}_{ij}(x-y) = \delta_{ij}\,\delta(x-y)
   + \Gamma_{\text{t}}\chi^0_{ij}(x-y)\quad,
\label{eq:2.10a}
\ee
where
\be
\chi^0_{ij}(x-y) = \frac{1}{2}\,
   \tr\left(\tilde{G}(y-x)\gamma_i\tilde{G}(x-y)\gamma_j\right)\quad.
\label{eq:2.10b}
\ee
\ese
In Appendix \ref{app:B} we show that $\chi$ is directly proportional to the
magnetic spin susceptibility $\chi_{\text{s}}$ in a Gaussian approximation. The
matrix $\Gamma$, which determines both the $\delta\Lambda$ and the
$\delta\mathcal{G}$ propagators, is given by
\bse
\label{eqs:2.11}
\be
\Gamma = \Gamma^{(1)} + \Gamma^{(2)}\quad,
\label{eq:2.11a}
\ee where
\bea
\Gamma^{(1)}_{\alpha\beta,\alpha'\beta'}(xy,x'y')&=&\frac{1}{2}\,
   \tilde{G}_{\alpha\beta'}(x-y')\,\tilde{G}_{\alpha'\beta}(x'-y)\quad,
\nonumber\\
\label{eq:2.11b}\\
\Gamma^{(2)}_{\alpha\beta,\alpha'\beta'}(xy,x'y')&=&\frac{\Gamma_{\text{t}}}{4}
   \int dz\,dz' \sum_{ij}\sum_{\gamma,\delta,\gamma',\delta'}
   \tilde{G}_{\alpha\gamma}(x-z)
\nonumber\\
&&\hskip -70pt \times \left(\gamma_i\right)_{\gamma\delta}
   \tilde{G}_{\delta\beta}(z-y)\,\chi_{ij}(z-z')\,
   \tilde{G}_{\alpha'\gamma'}(x'-z')\,
\nonumber\\
&&\hskip -70pt \times \left(\gamma_j\right)_{\gamma'\delta'}\,
   \tilde{G}_{\delta'\beta'}(z'-y')\quad.
\label{eq:2.11c}
\eea
\ese%
Notice that the $\delta\mathcal{G}$ vertex in Eq.\ (\ref{eq:2.9}) is minus the
inverse of the $\delta\Lambda$ vertex. This property will be important in what
follows.

\begin{figure}[t]
\vskip 10mm
\includegraphics[width=8cm]{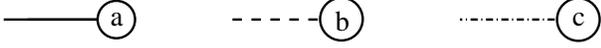}
\caption{\label{fig:2} Contributions to the linear part of the action,
  $\mathcal{A}^{(1)}$. Solid lines denote $\delta\bm{M}$, dashed lines
  $\delta\Lambda$, and dashed-dotted lines $\delta\mathcal{G}$.}
\end{figure}
We now consider the non-Gaussian terms in the action, which also are affected
by the shifts that diagonalize $\mathcal{A}^{(2)}$. The terms linear in the
fluctuations are represented graphically in Fig.\ \ref{fig:2}. Explicitly, we
find
\bse
\label{eqs:2.12}
\be
\mathcal{A}^{(1)} = \left({\bm a}\left\vert\right.\delta{\bm{M}}\right)
                    + \left(b\left\vert\right.\delta\Lambda\right)
                    + \left(c\left\vert\right.\delta\mathcal{G}^{\text{T}}
                           \right)\quad,
\label{eq:2.12a}
\ee
with
\be
a_i = \sqrt{2\Gamma_{\text{t}}}\,\left[-m\delta_{i3} + \frac{T}{2V}\,
      \Tr \left(\tilde{G}\gamma_i\right)\right]\quad,
\label{eq:2.12b}
\ee
and
\begin{widetext}
\bea
b_{\alpha\beta}(x,y)&=&\langle\mathcal{G}^{\text{T}}\rangle_{\alpha\beta}(x-y)
   - \frac{1}{2}\,\tilde{G}_{\alpha\beta}(x-y)
   +\sqrt{2\Gamma_{\text{t}}}\int dx'dy'\sum_{ij}\sum_{\gamma\delta}
     \tilde{G}_{\alpha\gamma}(x-x')\,\left(\gamma_i\right)_{\gamma\delta}
      \tilde{G}_{\delta\beta}(x'-y)\,\chi_{ij}(x'-y')\,a_j\ ,
\nonumber\\
\label{eq:2.12c}\\
c_{\alpha\beta}(x,y)&=&\lambda_{\alpha\beta}(x-y)
   + \int dx'dy' \sum_{\alpha'\beta'} b_{\alpha'\beta'}(x',y')\,
        \Gamma^{-1}_{\alpha'\beta',\alpha\beta}(x'y',xy)\quad.
\label{eq:2.12d}
\eea
\end{widetext}
\ese%

\begin{figure}[t]
\includegraphics[width=8cm]{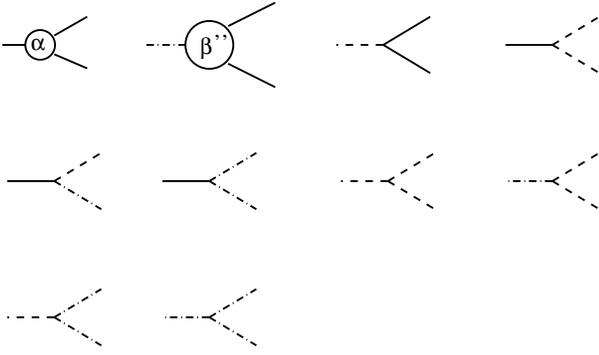}
\caption{\label{fig:3} Contributions to the cubic part of the action,
  $\mathcal{A}^{(3)}$.}
\end{figure}

For the cubic terms we find
\begin{widetext}
\bse
\label{eqs:2.13}
\bea
\mathcal{A}^{(3)}&=&\int dx\,dy\,dz\sum_{ijk}\alpha_{ijk}(x,y,z)\,
   \delta M_i(x)\,\delta M_j(y)\,\delta M_k(z)
   + \int dx\,dx'\,dx''\,dy''\sum_{ij}\sum_{\alpha\beta}
     {\beta''_{\alpha\beta}}^{ij} (x,x';x'',y'')\,
       \delta M_i(x)\,
\nonumber\\
&&\hskip 200pt \times \delta M_j(x')\,
        \delta\mathcal{G}^{\text{T}}_{\alpha\beta} (x'',y'')
       + (\text{other}\ \text{terms})\quad,
\label{eq:2.13a}
\eea
The vertices read
\be
\alpha_{ijk}(x,y,z) = \frac{1}{6}\,(2\Gamma_{\text{t}})^{(3/2)}\,
   \tr \left[\tilde{G}(z-x)\,\gamma_i\,\tilde{G}(x-y)\,\gamma_j\,
       \tilde{G}(y-z)\,\gamma_k\right]\quad,
\label{eq:2.13b}
\ee
and
\be
{\beta''_{\alpha\beta}}^{ij} (x,x';x'',y'') = \int dz\,dz'\sum_{\alpha'\beta'}
   {\beta'_{\alpha'\beta'}}^{ij}(x,x';zz')\,
   \Gamma^{-1}_{\alpha'\beta',\alpha\beta}(zz',x''y'')\quad,
\label{eq:2.13c}
\ee
where
\bea
{\beta'_{\alpha\beta}}^{ij} (x,x';x'',y'')&=&
   {\beta_{\alpha\beta}}^{ij} (x,x';x'',y'')
   + \frac{3}{2}\,\sqrt{\Gamma_{\text{t}}}\int dz\,dz'
     \sum_{i'j'}
     \sum_{\alpha'\beta'} \alpha_{ijj'}(x,x',z)\,
     \left(\gamma_{i'}\right)_{\beta'\alpha'}\,\chi_{i'j'}(z'-z)\,
     \tilde{G}_{\alpha'\beta}(z'-y'')\,
\nonumber\\
&&\hskip 250pt \times
     \tilde{G}_{\alpha\beta'}(x''-z') \quad,
\label{eq:2.13d}
\eea
with
\be
{\beta_{\alpha\beta}}^{ij} (x,x';x'',y'') = -\Gamma_{\text{t}}\,
   \left[\tilde{G}(x''-x)\,\gamma_i\,\tilde{G}(x-x')\,\gamma_j\,
 \tilde{G}(x'-y'')\right]_{\alpha\beta}\quad.
\label{eq:2.13e}
\ee
\ese%
\end{widetext}
We note that $\beta$ is the original $(\delta {\bm M})^2\delta\Lambda$ vertex.
Via the shifts of $\delta {\bm M}$ and $\delta\Lambda$ that diagonalize
${\mathcal A}^{(2)}$ it generates the $(\delta {\bm M})^2\delta\mathcal{G}$
vertex $\beta''$. The `other terms' in Eq.\ (\ref{eq:2.13a}) we will not need
explicitly. Their structure is displayed in Fig. \ref{fig:3}.

We are now in a position to determine the magnetic and superconducting
equations of state. In I this was done by means of integrating out the
fluctuations and minimizing the free energy with respect to the order
parameters. This procedure is not straightforward,\cite{errors_footnote} and it
becomes more confusing the more order-parameter fields one needs to consider.
For our present purposes, where we have one magnetic and two superconducting
order parameters, we therefore prefer a generalization of Ma's
method.\cite{Ma_1976}

\subsection{The magnetic equation of state}
\label{subsec:II.C}

\begin{figure}[tb]
\includegraphics[width=8cm]{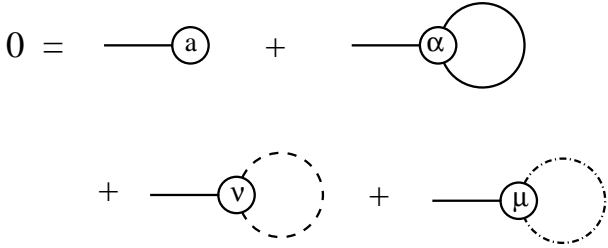}
\caption{\label{fig:4} The magnetic equation of state to one-loop order.}
\end{figure}

Following Ma,\cite{Ma_1976} we determine the magnetic equation of state by
requiring
\be
\langle\delta\bm{M}(\bm{x})\rangle = 0 \quad.
\label{eq:2.14}
\ee
This is a formal expression for the exact equation of state. By expanding the
action in powers of the fluctuations, as we have done above, it can be
evaluated order by order in a loop expansion. To one-loop order, we obtain the
diagrams shown in Fig.\ \ref{fig:4}.
Consider the last two diagrams. The original action, Eq.\ (\ref{eq:2.1}),
contained no $\delta\mathcal{G}$ legs at cubic order; these were produced by
the shift
\be
\delta\Lambda \rightarrow \Gamma^{-1}\delta\mathcal{G}^{\text{T}}\quad,
\label{eq:2.15}
\ee
that decouples $\delta\Lambda$ and $\delta\mathcal{G}$. As a result, the
vertices in these two diagrams, which in Fig.\ \ref{fig:4} are denoted by $\nu$
and $\mu$, respectively, are multiplicatively related by two factors of
$\Gamma^{-1}$. Symbolically,
\begin{equation*}
\nu = \mu\,\Gamma^{-2}\quad.
\end{equation*}
Together with the relation between the $\delta\Lambda$ and $\delta\mathcal{G}$
propagators that was mentioned after Eq.\ (\ref{eq:2.11c}), this implies that
the last two diagrams cancel each other:
\begin{equation*}
\delta\bm{M}\,\nu\Gamma^{-1} - \delta\bm{M}\,\mu\Gamma = 0\quad.
\end{equation*}
Clearly, this mechanism is not restricted to these particular
diagrams, but yields the following general diagram
rule\cite{diagram_rule_footnote}

\smallskip\noindent
{\it Rule:} $\delta\Lambda$ loops and $\delta\mathcal{G}$ loops cancel
                    each other.

\smallskip
\noindent This is the reason why we did not need to explicitly determine the
`other terms' in Eq.\ (\ref{eq:2.13a}). An evaluation of the first two diagrams
in Fig.\ \ref{fig:4} yields
\bea
m\,\delta_{i3}&=&\frac{T}{2V} \Tr\left(\gamma_i\tilde{G}\right)
   + \frac{\Gamma_{\text{t}}}{2} \int dx\,dy \sum_{jk}\chi_{jk}(x-y)
\nonumber\\
&&\tr\left[\gamma_i\tilde{G}(-x)\gamma_j\tilde{G}(x-y)\gamma_k\tilde{G}(y)
     \right]\quad.
\label{eq:2.16}
\eea
The first term on the right-hand side represents the mean-field magnetic
equation of state, which was derived (by a different method) and discussed in
I,\cite{missing_factor_footnote} while the second term represents one-loop
fluctuation corrections.

\subsection{The superconducting equation of state}
\label{subsec:II.D}

\begin{figure}[tb]
\includegraphics[width=8cm]{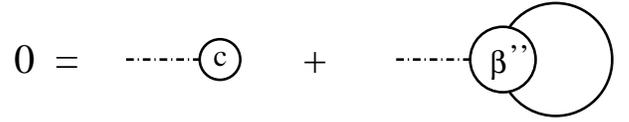}
\caption{\label{fig:5} The superconducting equation of state to one-loop
                       order.}
\end{figure}

We now determine the superconducting equation of state by requiring
\be
\langle\delta\mathcal{G}^{\text{T}}(x,y)\rangle = 0\quad.
\label{eq:2.17}
\ee
Taking into account the diagram rule from the preceding subsection, this
condition is graphically represented in Fig.\ \ref{fig:5}.
A calculation yields
\bse
\label{eqs:2.18}
\bea
\lambda_{\alpha\beta}(x-y)&=& -\int dx'\,dy'\sum_{\alpha'\beta'}
   \Biggl[\,\delta_{\alpha'\beta'}(x'-y')
\nonumber\\
&&\hskip -50pt\left. + \frac{1}{2} \int dx''\,dy''\sum_{ij} \chi_{ij}(x'-y')\,
     \beta_{\alpha'\beta'}^{ij}(x'y',x''y'')\right]\,
\nonumber\\
&&\hskip 30pt \times
   \Gamma^{-1}_{\alpha'\beta',\alpha\beta}(x''y'',xy)\quad.
\label{eq:2.18a}
\eea
with $\chi$ from Eq.\ (\ref{eq:2.10a}), $\beta$ from Eq.\ (\ref{eq:2.13e}),
$\Gamma$ from Eqs.\ (\ref{eqs:2.11}), and
\be
\delta = \langle\mathcal{G}^{\text{T}}\rangle - \frac{1}{2}\,\tilde{G}\quad.
\label{eq:2.18b}
\ee
\ese%
Here we have used the magnetic equation of state, Eq.\ (\ref{eq:2.16}).

The equations (\ref{eqs:2.18}) for $\lambda$ must be supplemented by a relation
between $\tilde{G}$ and $\langle\mathcal{G}\rangle$. We stress again that
$\lambda \neq \langle\Lambda\rangle$, so one must {\em not} require
$\langle\delta\Lambda\rangle = 0$.\cite{method_footnote} Rather, we calculate
$\langle\mathcal{G}\rangle$ directly. Going back to the underlying fermionic
formulation of the action, it is easy to show that
\be
\langle\mathcal{G}^{\text{T}}\rangle(x,y) = \frac{1}{2}\,
   \langle\tilde{G}_{\bm{M},\Lambda}(x,y)\rangle\quad,
\label{eq:2.19}
\ee
where $\tilde{G}_{\bm{M},\Lambda}$ is $\tilde{G}$, Eq.\ (\ref{eq:2.7}),
but with $m$ and $\lambda$ replaced by the full fluctuating fields
$\bm{M}$ and $\Lambda$. In evaluating
$\langle\tilde{G}_{\bm{M},\Lambda}(x,y)\rangle$, we restrict ourselves
to one-loop order, and also to linear order in the magnetic susceptibility
$\chi$, as contributions quadratic in $\chi$ are indistinguishable from
two-loop contributions. We find
\be
\langle\mathcal{G}^{\text{T}}\rangle = \frac{1}{2}\,\tilde{G}
   + O(\chi^2,2-\text{loop})\quad.
\label{eq:2.20}
\ee
The first term in brackets in Eq.\ (\ref{eq:2.18a}) thus vanishes, and the
second term we again evaluate to linear order in $\chi$. We finally obtain the
superconducting equation of state in the form
\be
\lambda_{\alpha\beta}(x-y) = \Gamma_{\text{t}}\sum_{ij}\chi_{ij}(y-x)\,
   \left[\gamma_i\,\tilde{G}(y-x)\,\gamma_j\right]_{\beta\alpha}\quad.
\label{eq:2.21}
\ee

For later reference, we note the following. We are analyzing the equation of
state in a loop expansion, and our treatment constitutes a systematic one-loop
evaluation. However, since the superconductivity does not occur at all unless
one goes to one-loop order, while the magnetism appears already at zero-loop
order, the resulting equations of state are not on the same level physically.
Specifically, the magnetic equation of state, Eq.\ (\ref{eq:2.16}), contains
fluctuation effects, while the superconducting one, Eq.\ (\ref{eq:2.21}), does
not, except for the magnetic fluctuations that cause the superconductivity in
the first place. In fact, the generalized Eliashberg equations that follow from
Eq.\ (\ref{eq:2.21}) are analogous to conventional Eliashberg theory, which
neglects all superconducting fluctuations.

\subsection{The Eliashberg equations}
\label{subsec:II.E}

Writing Eq.\ (\ref{eq:2.21}) explicitly yields the desired Eliashberg equations
for superconductivity induced by magnetic fluctuations. By using Eqs.\
(\ref{eq:2.7}) and (\ref{eq:2.3}) we obtain a set of coupled equations for the
matrix elements of $\lambda$, Eqs.\ (\ref{eqs:2.6}),
\bse
\label{eqs:2.22}
\bea
\Delta_{\uparrow}(k)&=&\Gamma_{\text{t}}\int_q \chi_{\text{L}}(q-k)\,
   \Delta_{\uparrow}(q)/d_{\uparrow}(q)\quad,
\label{eq:2.22a}\\
\Delta_{\downarrow}(k)&=&\Gamma_{\text{t}}\int_q \chi_{\text{L}}(q-k)\,
   \Delta_{\downarrow}(q)/d_{\downarrow}(q)\quad,
\label{eq:2.22b}\\
\Sigma_{\uparrow}(k)&=&\Gamma_{\text{t}}\int_q \chi_{\text{L}}(q-k)\,
   G_{\uparrow}^{-1}(q)/d_{\uparrow}(q)
\nonumber\\
&&\hskip -35pt +2\Gamma_{\text{t}}\int_q \left[\chi_{\text{T},+}(q-k) +
   i\chi_{\text{T},-}(q-k)\right]\,G_{\downarrow}^{-1}(q)/d_{\downarrow}(q)\ ,
\nonumber\\
\label{eq:2.22c}\\
\Sigma_{\downarrow}(k)&=&\Gamma_{\text{t}}\int_q \chi_{\text{L}}(q-k)\,
   G_{\downarrow}^{-1}(q)/d_{\downarrow}(q)
\nonumber\\
&&\hskip -35pt +2\Gamma_{\text{t}}\int_q \left[\chi_{\text{T},+}(q-k) +
   i\chi_{\text{T},-}(q-k)\right]\,G_{\uparrow}^{-1}(q)/d_{\uparrow}(q)\ .
\nonumber\\
\label{eq:2.22d}
\eea
\ese%
The $\Delta^+$ obey the same equation as the $\Delta$. Here we have performed a
Fourier transform to fermionic Matsubara frequencies $\omega_n = 2\pi T
(n+1/2)$ and wave vectors ${\bm k}$, and we use the notation
$k\equiv(i\omega_n,\bm{k})$, $\int_k \equiv T\sum_n (1/V)\sum_{\bm{k}}$. We
have introduced
\be
d_{\sigma}(k) = G_{\sigma}^{-1}(k)\,G_{\sigma}^{-1}(-k) + \Delta_{\sigma}(k)\,
                \Delta^{+}_{\sigma}(k)\quad,
\label{eq:2.23}
\ee
and the $G_{\uparrow,\downarrow}^{-1}$ are the inverse `normal' Green functions
(see Eqs.\ (\ref{eq:2.20}), (\ref{eq:2.7}), and (\ref{eq:2.4a}, \ref{eq:2.4b}),
\be
G_{\sigma}^{-1}(k) = i\omega_n - \xi_{\bm{k},\sigma} - \Sigma_{\sigma}(k)\quad,
\label{eq:2.24}
\ee
with $\xi_{{\bm k},\sigma} = {\bm k}^2/2m_{\text{e}} - \mu - \sigma\delta$.
Here $\sigma = \uparrow,\downarrow \equiv\pm$, and $\delta =
\Gamma_{\text{t}}\,m$ is the Stoner gap or exchange splitting. Finally, we have
used the fact that the magnetic susceptibility tensor $\chi$ in the presence of
a magnetization has the structure\cite{Forster_1975}
\bea
\chi = \begin{pmatrix} \chi_{\text{T},+} & \chi_{\text{T},-} & 0 \\
                     - \chi_{\text{T},-} & \chi_{\text{T},+} & 0 \\
                     0 & 0 & \chi_{\text{L}}
                    \end{pmatrix}\quad.
\label{eq:2.25}
\eea
This structure holds in general, and in particular for the explicit approximate
expression for $\chi$ given by Eqs.\ (\ref{eqs:2.10}). In a superconducting
phase, with non-vanishing gap functions, $\chi$ will depend on the gap. This
gives rise to a complicated feedback mechanism that is characteristic of any
purely electronic mechanism for superconductivity.\cite{Pao_Bickers_1991} We
will discuss this feedback in the following sections.

\section{The magnetic susceptibility in the presence of superconductivity}
\label{sec:III}

The Eliashberg equations (\ref{eqs:2.22}) require the magnetic susceptibility
$\chi$ as input, just like the Eliashberg equations for conventional
superconductivity require the phonon propagator as input. There are two
possible attitudes one can take at this point. In principle, one could use
experimental results for $\chi$, in analogy to experimental phonon spectra
being used as input for solving the conventional Eliashberg equations.
Sufficiently detailed information for $\chi$, however, is not available.
Alternatively, one can calculate or model $\chi$ in an effort to construct a
self-contained theory. This is what we will do now. For the purpose of
determining the phase diagram for up-spin and down-spin superconductivity, we
need $\chi$ in two phases. The up-spin superconducting $T_{\text{c}}$ is
determined by $\chi$ in the normal conducting ferromagnetic phase, NCFM in
Fig.\ \ref{fig:1}. This was discussed in I, and we briefly recall the result in
Sec.\ \ref{subsec:III.A}. For the down-spin superconducting $T_{\text{c}}$, we
need $\chi$ in the phase that has both magnetic and up-spin superconducting
order, SCFM I in Fig.\ \ref{fig:1}. This is discussed in Sec.\
\ref{subsec:III.B}.

\subsection{Normal conducting ferromagnetic phase}
\label{subsec:III.A}

The Gaussian theory, Sec.\ \ref{subsec:II.B}, yields an explicit expression for
$\chi$, viz., Eqs.\ (\ref{eqs:2.10}). For the normal conducting ferromagnetic
phase, this was evaluated in I. For the transverse susceptibility tensor at
small wave vectors and frequencies, the result is
\bse
\label{eqs:3.1}
\bea
\chi_{\rm T,+}(k) &=& \frac{\delta/4\epsilon_{\rm F}}
                                             {1 - t}\,
   \left(\frac{1}{i\Omega_n/4\epsilon_{\rm F} + (\delta/2\epsilon_{\rm F})
                  b_{\rm T}({\bf k}/2k_{\rm F})^2}   \right.
\nonumber\\
&& \left. - \frac{1}{i\Omega_n/4\epsilon_{\rm F} - (\delta/2\epsilon_{\rm F})
                  b_{\rm T}({\bf k}/2k_{\rm F})^2}   \right)\quad,
\nonumber\\
\label{eq:3.1a}\\
\chi_{\rm T,-}(k)\hskip -1pt &=& \hskip -1pt
              \frac{-i\delta/4\epsilon_{\rm F}}{1 - t}\hskip -2pt
   \left(\hskip -1pt\frac{1}{i\Omega_n/4\epsilon_{\rm F}
        + (\delta/2\epsilon_{\rm F}) b_{\rm T}({\bf k}/2k_{\rm F})^2}   \right.
\nonumber\\
&& \left. + \frac{1}{i\Omega_n/4\epsilon_{\rm F} - (\delta/2\epsilon_{\rm F})
                  b_{\rm T}({\bf k}/2k_{\rm F})^2}   \right)\quad.
\nonumber\\
\label{eq:3.1b}
\eea
\ese
Here $k = (i\Omega_n,{\bm k})$ with $\Omega_n = 2\pi Tn$ a bosonic Matsubara
frequency. $t = 1 - 2N_{\text{F}}\Gamma_{\text{t}}$ is the mean-field distance
from the magnetic critical point, with $N_{\text{F}}$ the density of states per
spin at the Fermi surface, and $\epsilon_{\text{F}}$ is the Fermi energy.
$b_{\text{T}} = 1/3$ in our free-electron approximation, but more generally it
is a number of order unity. This result displays the magnons, or magnetic
Goldstone modes, that are a consequence of the spontaneously broken spin
rotation symmetry in a ferromagnetic phase. It provides a qualitatively correct
expression for the transverse spin susceptibility in such a phase.

For the longitudinal susceptibility, the Gaussian approximation yields
\bse
\label{eqs:3.2}
\be
\chi_{\rm L}({\bf k},i0) = \frac{1-t}{a_{\rm L}\vert t\vert
                         + b_{\rm L}({\bf k}/2k_{\rm F})^2}\quad,
\label{eq:3.2a}
\ee
with $a_{\text{L}}$ and $b_{\text{L}}$ constants of $O(1)$. Popular model
calculations give $a_{\text{L}} = 5/4$ and $b_{\text{L}} =
1/3$.\cite{Brinkman_Engelsberg_1968} In contrast to the transverse channel,
however, Eq.\ (\ref{eq:3.2a}) is {\em not} qualitatively correct. The reason is
the mode-mode coupling effect that couples $\chi_{\text{L}}$ to
$\chi_{\text{T}}$ and leads to a longitudinal susceptibility that diverges at
${\bm k}=0$ everywhere in the ferromagnetic phase.\cite{Brezin_Wallace_1973} As
was discussed in I, the one-loop expression
\bea
\chi_{\rm L}^{(1)}(k) &=& \frac{2\Gamma_{\rm t}}{\delta^2}
   \,\chi_{\rm L}(k)\int_q \left[\chi_{\rm T,+}(k-q)\,
                                        \chi_{\rm T,+}(q)\right.
\nonumber\\
&&+ \left. \chi_{\rm T,-}(k-q)\,\chi_{\rm T,-}(q)\right]\,
                      \chi_{\rm L}(k)\quad,
\label{eq:3.2b}
\eea
takes this effect adequately into account. The one-loop approximation
\be
\chi_{\text{L}}(k) = \chi_{\text{L}}^{(0)}(k) + \chi_{\text{L}}^{(1)}(k)\quad,
\label{eq:3.2c}
\ee
\ese
thus correctly reflects the behavior of $\chi_{\text{L}}$ at small wave numbers
and frequencies.

\subsection{Superconducting ferromagnetic phase}
\label{subsec:III.B}

We now consider the magnetic susceptibility $\chi$ in the ferromagnetic phase
with $\Delta_{\uparrow} \neq 0$, $\Delta_{\downarrow} = 0$, SCFM I in Fig.\
\ref{fig:1}. We first consider the Gaussian approximation defined by Eqs.\
(\ref{eqs:2.10}), and discuss the validity of this approximation later. In
terms of Green functions, the five nonzero matrix elements of $\chi_{ij}^{-1}$
read,
\bse
\label{eqs:3.3}
\bea
\chi_{33}^{-1}(k) &=& 1 + \frac{1}{2}\,\Gamma_{\text{t}}\int_q\Bigl\{\Bigl[
   {\tilde G}_{11}(q+k)\,{\tilde G}_{11}(q)
\nonumber\\
   &&\hskip -0pt + {\tilde G}_{22}(q+k)\,{\tilde G}_{22}(q)
   - {\tilde G}_{13}(q+k)\,{\tilde G}_{31}(q)\Bigr]
\nonumber\\
   && \hskip 60pt + (k\rightarrow -k) \Bigr\}\quad,
\label{eq:3.3a}\\
\chi_{22}^{-1}(k) &=& 1 + \Gamma_{\text{t}} \int_q {\tilde G}_{11}(q)\,
   \left[{\tilde G}_{22}(q+k) + {\tilde G}_{22}(q-k)\right]
\nonumber\\
   &=& \chi_{11}^{-1}(k) \quad,
\label{eq:3.3b}\\
\chi_{12}^{-1}(k) &=& i\Gamma_{\text{t}}\int_q {\tilde G}_{11}(q)
             \,\left[{\tilde G}_{22}(q+k) - {\tilde G}_{22}(q-k)\right]
\nonumber\\
&=& -\chi_{21}^{-1}(k) \quad.
\label{eq:3.3c}
\eea
\ese
We need the susceptibility only at zero external frequency, and we perform the
integrals in an approximation that neglects the normal self energy as well as
the frequency dependence of $\Delta_{\uparrow}$. That is, we approximate
$\Delta_{\uparrow}(k) \approx \Delta_{\uparrow}\,{\hat k}_z$, with ${\hat k}_z$
the $z$-component of the unit wave vector. It is convenient to do the summation
over frequencies first, and to replace the wave number summation by an integral
over $\xi_{\bm q}$. The calculations are very similar to those of
susceptibilities in an s-wave superconductor.\cite{us_sc_response}

One readily finds that, as in the case of s-wave superconductors,
$\chi_{33}^{-1}({\bm k},i\Omega = 0)$ is independent of the gap. For
$\chi_{\text{L}} = 1/\chi_{33}^{-1}$ in Gaussian approximation, Eq.\
(\ref{eq:3.2a}) is therefore still valid. In the transverse channel the
situation is more complicated. It is illustrative to first consider the case of
zero external frequency and wave number. From Eq.\ (\ref{eq:3.3b}), and using
the magnetic equation of state, Eq.\ (\ref{eq:2.16}), in zero-loop
approximation (i.e., neglecting the second term on the right-hand side), we
have
\bea
\chi_{22}^{-1}(k=0) &=& 1 + 2\Gamma_{\text{t}}\int_q {\tilde G}_{11}(q)\,
   {\tilde G}_{22}(q)
\nonumber\\
&&\hskip -50pt = -\,\frac{\Gamma_{\text{t}}}{\delta} \int_q
   \vert\Delta_{\uparrow}(q)\vert^2\
   \frac{G_{\downarrow}(q)}{\vert G_{\uparrow}^{-1}\vert^2
      + \vert\Delta_{\uparrow}(q)\vert^2}
       \quad.
\label{eq:3.4}
\eea
We see that, within the framework of our approximations, the transverse
susceptibility has a mass proportional to $\Delta_{\uparrow}^2$. For small
values of the frequency, the wave number, and $\Delta_{\uparrow}$, we thus
obtain, instead of Eqs.\ (\ref{eqs:3.1}),
\bse
\label{eqs:3.5}
\begin{widetext}
\bea
\chi_{\text{T},+}({\bm k},i\Omega) &=& \frac{\delta/4\epsilon_{\text{F}}}{1-t}\
   \left(\frac{1}{i\Omega/4\epsilon_{\text{F}} + (\delta/2\epsilon_{\text{F}})\,
   b_{\text{T}}\,\left[({\bm k}/2k_{\text{F}})^2 +
   (\Delta_{\uparrow}/2\delta)^2\,g(\delta/2T)\right]} - (\delta\rightarrow
   -\delta)\right)\quad,
\label{eq:3.5a}\\
\chi_{\text{T},-}({\bm k},i\Omega)&=&\frac{-i\delta/4\epsilon_{\text{F}}}{1-t}\
   \left(\frac{1}{i\Omega/4\epsilon_{\text{F}} + (\delta/2\epsilon_{\text{F}})\,
   b_{\text{T}}\,\left[({\bm k}/2k_{\text{F}})^2 +
   (\Delta_{\uparrow}/2\delta)^2\,g(\delta/2T)\right]} + (\delta\rightarrow
   -\delta)\right)\quad.
\label{eq:3.5b}
\eea
\end{widetext}
Here
\be
g(\delta/2T) = \frac{-6\delta}{N_{\text{F}}\vert\Delta_{\uparrow}\vert^2}
   \int_q \vert\Delta_{\uparrow}(q)\vert^2\,\vert G_{\uparrow}(q)\vert^2\,
   G_{\downarrow}(q)\quad.
\label{eq:3.5c}
\ee
\ese
In the limit $T\ll\epsilon_{\text{F}}$ we obtain
\be
g(y) = \int_0^{\infty} dx\ \frac{1}{x(1-x^2)}\,\tanh(xy)
     = \begin{cases} \ln y & \text{if $y\gg 1$}\quad,\\
                     a\,y^2& \text{if $y\ll 1$}\quad,
       \end{cases}
\label{eq:3.6}
\ee
where $a = 0.8525\ldots$.

The above results have been obtained in an approximation that neglects all
fluctuations of the superconducting order parameter, see the remark at the end
of Sec.\ \ref{subsec:II.D}). This raises the question whether the mass is
generic, or a result of our approximations. To answer this question we
consider, in Appendix \ref{app:A.2}, a LGW theory that treats magnetic and
superconducting fluctuations on equal footing. Both a general symmetry analysis
and an explicit calculation within the framework of the LGW theory show that
the result obtained above is correct for all wave numbers only in the limit
$\xi_{\Delta}/\xi_m \to \infty$, where $\xi_{\Delta}$ and $\xi_m$ is the
superconducting and magnetic (zero-temperature) coherence length, respectively.
For a finite value of this ratio, $\chi_{\text{T}}$ as a function of the wave
number displays a shoulder, but still diverges for asymptotically small values
of $\vert{\bm k}\vert$. We conclude that, first, the presence of
$\Delta_{\uparrow}$ qualitatively changes the dispersion relation of the
transverse magnons, and second, the Eqs.\ (\ref{eqs:3.5}) constitute an upper
bound on this change, which becomes exact in the limit $\xi_{\Delta}/\xi_m \to
\infty$. We will discuss this point further in Sec.\ \ref{sec:V} below.

Equations (\ref{eqs:3.2}) for the longitudinal susceptibility in one-loop
approximation remain valid. This expression, with $\chi_{\text{T},\pm}$ from
Eqs.\ (\ref{eqs:3.5}), and used in the Eliashberg equations, will provide an
upper limit for the effect of $\Delta_{\uparrow}$ on $T_{\text{c}\downarrow}$.

\section{The structure of the superconducting transition}
\label{sec:IV}

We now are in a position where we can solve the gap equations (\ref{eqs:2.22})
for the down-spin superconducting critical temperature,
$T_{\text{c}\downarrow}$. We will do so in an approximation that is analogous
to the one employed in I for $T_{\text{c}\uparrow}$.

\subsection{Linearized gap equation, and $T_{\text{c}}$ formula}
\label{subsec:IV.A}

We start by linearizing the gap equations in
$\Delta_{\downarrow}$,\cite{1st_order_footnote} while keeping the full
dependence on $\Delta_{\uparrow}$. Equations (\ref{eqs:2.22}) become
\bse
\label{eqs:4.1}
\bea
\Delta_{\downarrow}(k)&=&\Gamma_{\text{t}}\int_q \chi_{\text{L}}(q-k)\,
   \Delta_{\downarrow}(q)\,\vert G_{\downarrow}(q)\vert^2\quad,
\label{eq:4.1a}\\
\Sigma_{\downarrow}(k)&=&\Gamma_{\text{t}}\int_q \chi_{\text{L}}(q-k)\,
   G_{\downarrow}(-q)
\nonumber\\
&&\hskip -35pt +2\Gamma_{\text{t}}\int_q \left[\chi_{\text{T},+}(q-k) +
   i\chi_{\text{T},-}(q-k)\right]\,G_{\uparrow}^{-1}(q)/d_{\uparrow}(q)\ ,
\nonumber\\
\label{eq:4.1b}\\
\Delta_{\uparrow}(k)&=&\Gamma_{\text{t}}\int_q \chi_{\text{L}}(q-k)\,
   \Delta_{\uparrow}(q)/d_{\uparrow}(q)\quad,
\label{eq:4.1c}\\
\Sigma_{\uparrow}(k)&=&\Gamma_{\text{t}}\int_q \chi_{\text{L}}(q-k)\,
   G_{\uparrow}^{-1}(q)/d_{\uparrow}(q)
\nonumber\\
&&\hskip -35pt +2\Gamma_{\text{t}}\int_q \left[\chi_{\text{T},+}(q-k) +
   i\chi_{\text{T},-}(q-k)\right]\,G_{\downarrow}(-q)\quad,
\nonumber\\
\label{eq:4.1d}
\eea
\ese%
Notice that the down-spin and up-spin equations are coupled in two ways;
directly via the explicit dependence of the transverse spin-fluctuation
contribution to $\Sigma_{\downarrow}$ on $\Delta_{\uparrow}$, and indirectly
via the dependence of the $\chi_{\text{T}}$ on $\Delta_{\uparrow}$. The former
effect means that in the second contribution to $\Sigma_{\downarrow}$, the wave
number is not strictly pinned to the up-spin Fermi surface. However, this
effect is only on the order of $\Delta_{\uparrow}/\epsilon_{\text{F}}$, and we
neglect it. The up-spin and down-spin equations then formally decouple, except
for the dependence of $\chi$ on $\Delta_{\uparrow}$. We can then solve for
$T_{\text{c}\downarrow}$, in terms of $\chi$, in the same McMillan-Allen-Dynes
approximation that was employed in I for $T_{\text{c}\uparrow}$. We find
\be
T_{\text{c}\downarrow} = T_0(t)\,e^{-(1+d^0_{\text{L}\downarrow} +
             2d^0_{\text{T}\downarrow})/d^1_{\text{L}\downarrow}}\quad.
\label{eq:4.2}
\ee
Here $T_0(t)$ is the same temperature scale that was used in I,
viz.,\cite{Fay_Appel_1980}
\be
T_0(t) = T_0\ \left[\Theta(t)\,t + \Theta(-t)\,5\vert t\vert/4\right]\quad,
\label{eq:4.3}
\ee
with $T_0$ a microscopic temperature on the order of the Fermi temperature. The
coupling constants $d$ are given by
\bse
\label{eqs:4.4}
\bea
d^1_{\text{L}\downarrow} &=& \Gamma_{\text{t}}\,N_{\text{F}\downarrow} \int_0^2
dx\ x\,(1-x^2/2)\,\chi_{\text{L}}(xk_{\text{F}\downarrow},i0)\quad, \nonumber\\
\label{eq:4.4a}\\
d^0_{\text{L}\downarrow} &=& \Gamma_{\text{t}}\,N_{\text{F}\downarrow} \int_0^2
dx\ x\,\chi_{\text{L}}(xk_{\text{F}\downarrow},i0)\quad,
\label{eq:4.4b}\\
d^0_{\text{T}\downarrow} &=& \Gamma_{\text{t}}\,N_{\text{F}\downarrow}
\int_{y_-}^{y_+} dx\ x\,\chi_{\text{T},+}(xk_{\text{F}\downarrow},i0)\quad,
\label{eq:4.4c}
\eea
\ese
with $y_{\pm} = k_{\text{F}\uparrow}/k_{\text{F}\downarrow} \pm 1 =
\delta_+/\delta_- \pm 1$. Here $k_{\text{F}\sigma}$ and $N_{\text{F}\sigma}$,
$\sigma = \uparrow,\downarrow$, are the Fermi wave number and the density of
states at the up-spin and down-spin Fermi surface, respectively.

\subsection{The superconducting phase diagram}
\label{subsec:IV.B}

\begin{figure}[t]
\vskip 10mm
\includegraphics[width=95mm]{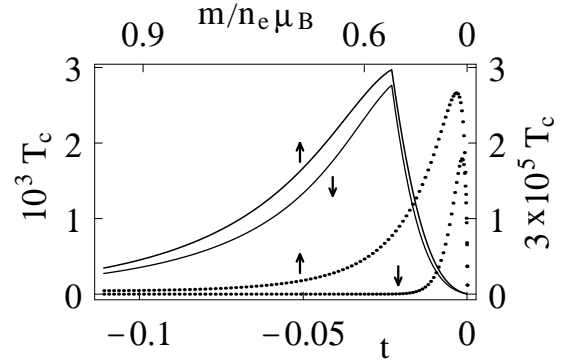}
\caption{\label{fig:6} $T_{\text{c}\uparrow}$ and $T_{\text{c}\downarrow}$ in
zero-loop (dotted lines, right scale) and one-loop (solid lines, left scale)
approximation. Parameter values used are $b_{\text{L}}=0.23$,
$b_{\text{T}}=0.25$.}
\end{figure}

For calculating $T_{\text{c}}$ numerically, we use the same mean-field relation
between $t$ and $m$ or $\delta$ as in I, namely,\cite{Brinkman_Engelsberg_1968}
\bse
\label{eqs:4.5}
\bea
t &=& 1 - (1 + 3\eta^2)^{1/3}/(1 + \eta^2/3)\quad,
\label{eq:4.5a}\\
m/\mu_{\text{B}}n_{\text{e}} &=& 3\eta(1 + \eta^2/3)/(1 + 3\eta^2)\quad,
\label{eq:4.5b}
\eea
\ese
with $n_{\text{e}}$ the electron number density and $\mu_{\text{B}}$ the Bohr
magneton. In zero-loop approximation, with $\chi$ given by Eqs.\
(\ref{eq:3.2a}, \ref{eqs:3.5}), $T_{\text{c}\downarrow}$ for a given
$\Delta_{\uparrow}$ is now easy to calculate. Rather than solving the up-spin
Eliashberg equations for $\Delta_{\uparrow}$, we have approximated
$\Delta_{\uparrow}\approx 2T_{\text{c}\uparrow}$. $T_{\text{c}\uparrow}$ is the
same as in I.\cite{numerical_mistake_footnote} Our results are very similar to
those obtained by Fay and Appel\cite{Fay_Appel_1980} and are shown in Fig.\
\ref{fig:6}. The effect of $\Delta_{\uparrow}$ on $T_{\text{c}\downarrow}$ is
so small that it is not visible on the scale of the figure. At the zero-loop
level, the feedback effect is thus unobservably small.

At the one-loop level, with $\chi_{\text{L}}$ given by Eqs.\ (\ref{eqs:3.2})
with Eqs.\ (\ref{eqs:3.5}) as input, the calculation is more complicated. We
have approximated the effect of the mass in $\chi_{\text{T}}$ by a lower cutoff
in the frequency summation, $\Omega_n^- =
2b_{\text{T}}\delta(\Delta_{\uparrow}/\delta)^2\ln(\delta/T)$. This allows to
perform the wave-number integral analytically, as was described in I. As in the
zero-loop calculation, we replace $\Delta_{\uparrow}$ by
$2T_{\text{c}\uparrow}$. This procedure will overestimate the effect of
$\Delta_{\uparrow}$ on $T_{\text{c}\downarrow}$. The calculation then proceeds
as for $T_{\text{c}\uparrow}$. A representative result is shown in Fig.\
\ref{fig:6}. We see that the relative difference between $T_{\text{c}\uparrow}$
and $T_{\text{c}\downarrow}$ in one-loop approximation is on the same order as
in zero-loop approximation, viz., at most about 10\%. The theory thus predicts
two superconducting transitions close to one another. In the notation of the
schematic phase diagram in Fig.\ \ref{fig:1} this means that the SCFM I phase
is very narrow.

\subsection{The specific heat}
\label{subsec:IV.C}

\begin{figure}[t]
\vskip 0mm
\includegraphics[width=6cm]{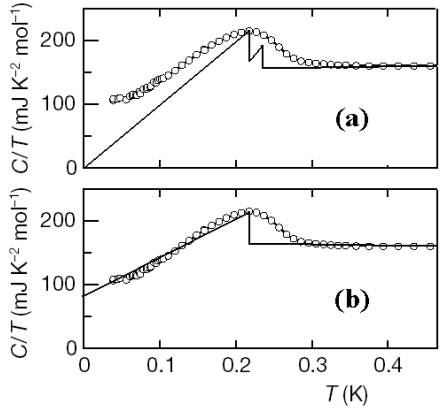}
\vskip -5mm
\caption{\label{fig:7} The specific heat coefficient of URhGe as observed in
  Ref.\ \protect\onlinecite{Aoki_et_al_2001} (circles). The solid line in (a)
  shows the qualitative theoretical prediction for a perfect sample based on the
  present analysis. The one in (b) would apply if the down-spin electrons
  remained unpaired to the lowest observable temperature. The size of the
  discontinuity is known theoretically only in the weak-coupling
  limit;\protect\cite{Vollhardt_Woelfle_1990} for the purpose of this qualitative
  illustration we have adjusted it to fit the maximum in the data.}
\end{figure}

The above results for the critical temperatures have implications for the
specific heat. The two superconducting transitions close to one another imply
two features in the specific heat. At the mean-field level, the specific heat
at a superconducting transition has a discontinuity, and for our choice of an
order parameter the low-temperature specific heat is a quadratic function of
the temperature.\cite{Vollhardt_Woelfle_1990} One therefore qualitatively
expects the specific heat coefficient as a function of temperature to behave as
shown in Fig.\ \ref{fig:7}(a). This should be compared to a situation where, as
suggested in Ref.\ \onlinecite{Aoki_et_al_2001}, the down-spin electrons do not
pair down to the lowest observable temperature, which would result in a
specific heat coefficient as shown qualitatively in Fig.\ \ref{fig:7}(b). The
data of Ref.\ \onlinecite{Aoki_et_al_2001} show a very broad feature, even for
the best samples, which is consistent with either scenario. The prediction of
the present theory is that, with further improving sample quality, the residual
value of the specific heat coefficient will decrease, and the broad peak will
be resolved into two discontinuities.

\section{Discussion and conclusion}
\label{sec:V}

Let us further discuss several aspects of our theory and its implications.
First, we consider in more detail our results regarding the magnetic
susceptibility. In a normal conducting ferromagnetic phase, the transverse spin
susceptibility is soft, displaying ferromagnetic magnons with a quadratic
dispersion relation. The scale in the dispersion relation is set by the Stoner
gap or exchange splitting $\delta$, which is proportional to the magnetization.
In natural units, measuring wave numbers in units of the magnetic coherence
length $\xi_m$, which is on the order of the Fermi wave number $k_{\text{F}}$,
we have
\begin{equation*}
\Omega \sim \delta\ ({\bf k}\xi_m)^2\quad.
\end{equation*}
At zero frequency, the dimensionless transverse magnetic susceptibility
diverges for ${\bm k} \to 0$ like
\begin{equation*}
\chi_{\text{T}}({\bm k}) = c/({\bm k}\xi_m)^2\quad.
\end{equation*}
For a given distance from the magnetic phase transition, $c$ is a number of
order one whose exact value depends on the details of the model. The Gaussian
approximation we have employed correctly reflects these exact features, see
Eqs.\ (\ref{eqs:3.1}). In the presence of an up-spin gap, the structure of the
magnon changes qualitatively. The Gaussian approximation for $\chi_{\text{T}}$
predicts a true gap, Eqs.\ (\ref{eqs:3.5}). This is a consequence of the fact
that the Gaussian approximation effectively treats the superconducting gap as a
fixed external field. The LGW theory of Appendix \ref{app:A} shows that this
does not correctly reflect the symmetry properties of the system, and that the
static $\chi_{\text{T}}$ actually has the structure
\bse
\label{eqs:5.1}
\be
\chi_{\text{T}}({\bm k}) = \frac{c}{({\bm k}\xi_m)^2}\ \cfrac{1}{1 + g\,\cfrac{
2N_{\text{F}}\Gamma_{\text{t}}\vert\Delta_{\uparrow}\vert^2/\delta^2} {({\bm
k}\xi_m)^2 + g\,(\xi_m/\xi_{\Delta})^2}}\quad.
\label{eq:5.1a}
\ee
Here $g$ is related to the coupling constant ${\tilde g}$ and the magnetic
order parameter ${\tilde m}$ of the LGW theory by $g = {\tilde g}\,{\tilde m}$.
This result shows that, in general, $\chi_{\text{T}}$ has a much more
complicated structure than in the absence of spin-triplet superconductivity. At
asymptotically small wave numbers, ${\bm k}^2 < g\,\xi_{\Delta}^{-2}$, $g$
drops out of the expression for $\chi_{\text{T}}$, and we have
\be
\chi_{\text{T}}({\bm k}) = \frac{c}{{\bm k}^2}\ \frac{1}{1 +
(\Delta_{\uparrow}/\delta)^2\,(\xi_{\Delta}/\xi_m)^2}\quad.
\label{eq:5.1b}
\ee
Restoring the frequency, the dispersion relation of the magnon then is
\begin{equation*}
\Omega \sim \delta\,\left[1 +
         (\Delta_{\uparrow}/\delta)^2\,(\xi_{\Delta}/\xi_m)^2\right]
         \,{\bm k}^2\quad.
\end{equation*}

We see that the magnon dispersion relation is substantially changed by the
presence of $\Delta_{\uparrow}$ as long as the condition expressed by Eq.\
(\ref{eq:A.10}) is fulfilled. Namely, the magnon is much stiffer, i.e., the
frequency rises much faster with the wave number, than in a normal conducting
ferromagnet. The magnitude of this effect will depend on the detailed
parameters of the material in question, but it clearly can be substantial:
Assuming a value of $\Delta_{\uparrow}$ on the order of the superconducting
$T_{\text{c}}$, or $\Delta_{\uparrow}\approx 1\text{K}$, a value of $\delta$ on
the order of 10 times the magnetic $T_{\text{c}}$,\cite{delta_footnote} or
$\delta\approx 100\text{K}$, and $\xi_{\Delta}/\xi_m\approx 1,000$, we find
that the prefactor in the dispersion relation is enhance by a factor of 100
over its value in the normal conducting phase.

For larger wave numbers, ${\bm k}^2 > g\,\xi_{\Delta}^{-2}$, we have
\be
\chi_{\text{T}}({\bm k}) \approx \frac{c}{({\bm k}\xi_m)^2 +
g\,2N_{\text{F}}\Gamma_{\text{t}}\,\vert\Delta_{\uparrow}\vert^2/\delta^2}\quad.
\label{eq:5.1c}
\ee
\ese
Using Eqs.\ (\ref{eqs:A.9}), we have the correspondence $g \cong g(\delta/2T)/3
\agt 1$, so we recover the result of the Gaussian theory, Eqs.\
(\ref{eqs:3.5}). As long as Eq.\ (\ref{eq:A.10}) is valid, $\chi_{\text{T}}$
thus displays a pronounced plateau as a function of the wave number, followed
by a very steep increase at asymptotically small wave numbers. The Gaussian
theory approximates this behavior by a true gap, ignoring the asymptotic
regime. In the limit $\xi_{\Delta}/\xi_m \to \infty$ the asymptotic regime
shrinks to zero, and the Gaussian theory becomes qualitatively correct.
However, in this context one should note that the LGW theory of Appendix
\ref{app:A} is valid only for wave numbers $\vert{\bm k}\vert <
\xi_{\Delta}^{-1}$. Within the LGW theory, the shoulder will lie in that regime
provided the coupling constant ${\tilde g}$ is sufficiently small. While the
above identification of ${\tilde g}$ with the parameters of the microscopic
theory makes ${\tilde g}$ effectively of order unity, given the initial sharp
rise of the magnon frequency with the wave number in the regime where the LGW
is valid, a pronounced shoulder in the dispersion relation at intermediate wave
numbers is inevitable. The wave number region where one expects this shoulder
is given by
\be
g\,\xi_{\Delta}^{-2} < {\bm k}^2 <
g\,(\vert\Delta_{\uparrow}\vert/\delta)^2\,\xi_m^{-2}\quad.
\label{eq:5.2}
\ee
If we use the same numbers as above, and $g=1$, $\xi_m = 1$\AA, the upper limit
of this wave-number range is given by $\vert{\bm k}\vert = 0.01$\AA. This is a
factor of 3 below the smallest wave numbers currently observable with neutron
scattering.\cite{Kittel_1996} However, in materials with smaller values of the
exchange splitting $\delta$ the plateau should be in an observable regime.

Second, we add some comments about our prediction of two superconducting
transitions that are separated only by a small temperature interval. Within our
model, we have found this prediction to be very robust, especially given that
all of our approximations have a tendency to overestimate the suppression of
$T_{\text{c}\downarrow}$ compared to $T_{\text{c}\uparrow}$. Even an
(artificial) increase of the gap in the transverse susceptibility by a factor
of 10 does not make a visible change in Fig.\ \ref{fig:6}. This reflects the
fact that the up-spin and down-spin pairing are mediated by the same effective
potential, viz., $\chi_{\text{L}}$, see Eq.\ (\ref{eq:4.4a}) and the
corresponding Eq.\ (3.14a) in I. While $\chi_{\text{L}}$ is modified by
$\Delta_{\uparrow}$, the effect is not sufficiently large to lead to a
substantial separation of $T_{\text{c}\uparrow}$ and $T_{\text{c}\downarrow}$.
This in turn means that the lower value of $T_{\text{c}\downarrow}$ is
overwhelmingly due to the lower value of the density of states at the down-spin
Fermi level. In this context we need to keep in mind that we have used a
free-electron model with parabolic bands. A complicated band structure could
lead to a drastically reduced value of $N_{\text{F}\downarrow}$, which in turn
would lead to a much lower value of $T_{\text{c}\downarrow}$. If experiments on
samples of improved quality should fail to show two transitions, this would be
the most likely explanation. This would be of interest also with regard to
distinguishing between the two proposed explanations for why the
superconductivity is observed in the ferromagnetic phase only: Sandeman et
al.\cite{Sandeman_et_al_2003} have proposed a mechanism based on an intricate
structure of the density of states, while the explanation proposed in I is
based on properties of the magnetic susceptibility.

Third, we come back to our discussion of the specific heat. The experiment of
Ref.\ \onlinecite{Aoki_et_al_2001} shows that the observed superconductivity is
indeed a bulk effect. What is not clear {\it a priori} is the origin of the
large residual value of the specific heat coefficient. While the down-spin
electrons remaining unpaired, as was suggested in Ref.\
\onlinecite{Aoki_et_al_2001}, is a possibility, normal-conducting regions
within the sample would have the same effect and would also lead to the
observed smearing of the discontinuity in the specific heat. With increasing
sample quality, the discontinuity should become sharper, and a crucial question
will be whether the residual value drops correspondingly. Of course, the
emergence of the predicted double feature from the narrowing peak will be the
most direct test of our predictions regarding the specific heat. In this
context it is interesting to note that such a split transition, with two
closely spaced discontinuities in the specific heat, has been observed in
UPt$_3$,\cite{Fisher_et_al_1989,Hasselbach_et_al_1989} but only after a long
period of gradually increasing sample quality. (UPt$_3$ is not ferromagnetic,
though, and the split transition has a physical origin that is very different
from what we have discussed.) Even in the best samples that show the split
transition, however, there is a substantial residual specific heat coefficient,
the origin of which is not quite clear. In a ferromagnetic superconductor, one
also has to keep in mind that the ground state will not be homogeneous, due to
the formation of a spontaneous vortex state.\cite{Greenside_Blount_Varma_1981,
Ng_Varma_1997, Radzihovsky_et_al_2001} Normal electrons in the vortex cores are
a possible source of a residual specific heat coefficient. This effect has been
neglected in the current theory and will be pursued in a future publication.

Fourth, we briefly comment on the fact that the mass, or pseudo-mass, induced
in the transverse magnetic susceptibility by the superconductivity, is a
singular function of the magnetization. This can be seen in Eqs.\
(\ref{eqs:3.5}), and it also leads to the factor of $1/\delta$ in the relation
between the phenomenological coupling constant ${\tilde g}$ in the LGW theory
and the microscopic parameters, Eq.\ (\ref{eq:A.9c}). This behavior can be
traced back to the behavior of the integral in Eq.\ (\ref{eq:3.5c}), and thus
ultimately to the Green functions and the soft particle-hole excitations that
are characteristic for itinerant electron systems. The singularity is therefore
a result of a coupling between the particle-hole excitations and the magnetic
and superconducting Goldstone modes. It is very similar in nature to, e.g., the
anomalous magnetization dependence of the magnon stiffness in a normal
conducting ferromagnetic phase that was discussed in Ref.\
\onlinecite{us_magnon_dispersion}.

In conclusion, we summarize our results. We have presented a consistent and
self-contained theory for the coexistence of superconductivity and
ferromagnetism in itinerant electron systems. We have presented a
field-theoretic formulation of this problem that allows for the determination
of both the magnetic and the superconducting equation of state in a systematic
loop expansion. This method, which utilizes Ma's procedure for generating
equations of state, remedies some shortcomings of the earlier theory presented
in I, which relied on a minimization of the free energy. The self-contained
character of the theory is achieved by means of explicit expressions for the
magnetic susceptibility, which is needed as input for the generalized
Eliashberg equations. These expressions have been evaluated in ferromagnetic
phases, both normal conducting and superconducting ones. This is a
generalization of the theory for the magnetic susceptibility that was developed
in I, and it explicitly takes into account the feedback effects that are
characteristic for any purely electronic mechanism for superconductivity.

We have explicitly evaluated this theory to one-loop order, and for a model
that allows for two components of the superconducting order parameter, one each
for Cooper pairs consisting of up-spin electrons and down-spin electrons,
respectively. The limitations of this one-loop approximation, which neglects
superconducting fluctuations and uses a zero-loop expression for the magnetic
susceptibility, have been discussed by means of a phenomenological LGW theory
that complements our microscopic theory. We have found that, for generic
parameter values, the two superconducting transitions that describe the pairing
of up-spin and down-spin electrons, respectively, occur close to one another,
with transition temperatures that typically differ by only on the order of
10\%. This suggests that, if the superconductivity observed in URhGe is indeed
of a spin-triplet p-wave type mediated by ferromagnetic fluctuations, then the
broad feature observed in the specific heat near the temperature of the
resistive transition\cite{Aoki_et_al_2001} should contain two transitions that
are close together. If samples of improved quality should show only one sharp
discontinuity in the specific heat, then this would be a strong argument
against the type of pairing we have assumed in this paper. A caveat is provided
by our assumption of parabolic bands, however, as discussed above. The presence
of spin-triplet superconductivity has further been shown to drastically change
the structure of the dispersion relation of the ferromagnetic magnons. In
materials with a small exchange splitting this effect is observable with
neutron scattering, and can also be used as a probe for the nature of the
superconductivity.

\acknowledgments We would like to thank Meigan Aronson, J. David Cohen, and
John Toner for helpful discussions. We are indebted to Thomas Vojta for
insisting that the question of Goldstone modes warranted a more detailed
discussion than we initially were prepared to give. This work was supported by
the NSF grant under Nos. DMR-01-32555 and DMR-01-32726.

\appendix

\section{LGW theory}
\label{app:A}

In this appendix we discuss the mean-field action $\mathcal{A}^{(0)}$, Eq.\
(\ref{eq:2.8}), or the corresponding Landau free energy density $f^{(0)} =
-(T/V)\mathcal{A}^{(0)}$, and its generalization to a phenomenological LGW
theory.

\subsection{Landau theory}
\label{app:A.1}

From Eq.\ (\ref{eq:2.8}) we have
\be
f^{(0)} = \frac{\Gamma_{\text{t}}}{2}\,m^2
          - \frac{T}{V}\,\Tr (\lambda\langle\mathcal{G}\rangle)
          - \frac{T}{2V}\,\Tr \ln\tilde{G}^{-1}\quad.
\label{eq:A.1}
\ee
In what follows we neglect the normal self-energy contribution to the
matrix $\lambda$, Eqs.\ (\ref{eqs:2.6}), which lead only to a trivial
renormalization of the normal Green function. An expansion in powers
of the order parameters $m$ and $\lambda$ then yields
\bea
f^{(0)} &=& \Tr\left(\lambda^{\text{T}}(\langle\mathcal{G}^{\text{T}}\rangle
                     - \tilde{G}_0/2)\right)
           - \frac{T}{4V}\,\Tr\left(\lambda^{\text{T}}\tilde{G}_0\right)^2
\nonumber\\
&& \hskip -40pt +  \frac{\Gamma_{\text{t}}}{2}\,t\,m^2
     - \frac{T\Gamma_{\text{t}}}{2V}\,m\,\Tr\left(\tilde{G}_0\,\gamma_3\,
                \tilde{G}_0\lambda^{\text{T}}\,\tilde{G}_0\lambda^{\text{T}}
                   \right) + \ldots
\label{eq:A.2}
\eea
Here $t = 1 - 2N_{\text{F}}\Gamma_{\text{t}}$ is the mean-field distance
from the magnetic critical point, and we have omitted cubic terms of
order $m^3$, $m^2\lambda$, and $\lambda^3$, as well as all quartic terms.

This mean-field free energy does not describe a superconducting transition;
this requires magnetic loops, as we have seen in the main text. The magnetic
part, however, agrees with an expansion of the mean-field magnetic
equation of state, Eq.\ (\ref{eq:2.14}). In particular,
the cubic term of $O(m\lambda^2)$ shown in Eq.\ (\ref{eq:A.2}) has
important consequences. For a nonzero superconducting order parameter
$\Delta_{\uparrow}$, Eq.\ (\ref{eqs:2.6}), the quantity
\bse
\label{eqs:A.3}
\be
h_{\text{eff}} = \frac{T\Gamma_{\text{t}}}{2V}\,\Tr\left(\tilde{G}_0\,\gamma_3\,
                \tilde{G}_0\lambda^{\text{T}}\,\tilde{G}_0\lambda^{\text{T}}
                   \right)
\label{eq:A.3a}
\ee
acts like an effective external magnetic field. Performing the trace, we have
\be
h_{\text{eff}} = -\Gamma_{\text{t}}\,\frac{T}{V}\sum_k \vert\Delta_{\uparrow}(k)
   \vert^2\,\vert\tilde{G}_0(k)\vert^2\,\tilde{G}_0(k)\quad.
\label{eq:A.3b}
\ee
\ese%
General arguments\cite{Ma_1976} show that such a term in the free energy leads
to a mass in the magnetic Goldstone mode, i.e., in the transverse spin
susceptibility,
\be
\chi_{\text{s}}^{-1}(k=0) = h_{\text{eff}}/m\quad.
\label{eq:A.4}
\ee
If we expand Eq.\ (\ref{eq:3.4}) in powers of $\Delta_{\uparrow}$ and $m$, we
see that this is indeed the same result we obtained in Sec.\ \ref{sec:III} by a
direct calculation of the susceptibility. The current derivation makes it
obvious that this result has been obtained while neglecting fluctuations. We
consider the influence of fluctuations in the next subsection.

\subsection{LGW theory}
\label{app:A.2}

The conclusion in the previous subsection, namely, that the transverse spin
susceptibility is massive, cannot be strictly correct. Even in the presence of
superconducting order, one expects the system to be invariant under rotations
of all spins, and this symmetry must manifests itself via a Goldstone mode to
which the magnetic susceptibility must couple. To investigate this point, we
consider the following phenomenological action,\cite{quartic_term_footnote}
\begin{widetext}
\bea
S[{\bm m},{\bm\phi}] &=& \int d{\bm x}\ \left[t_m\,{\bm m}^2({\bm x})
   + \xi_m^2\,\left(\nabla{\bm m}({\bm x})\right)^2 + u_m\,{\bm m}^4({\bm x})
   \right] + \int d{\bm x}\ \left[t_{\Delta}\,\vert{\bm\phi}({\bm x})\vert^2
   + \xi_{\Delta}^2\,\vert\nabla{\bm\phi}({\bm x})\vert^2
      + u_{\Delta}\,\vert{\bm\phi}({\bm x})\vert^4\right]
\nonumber\\
&&-i{\tilde g}\int d{\bm x}\ {\bm m}({\bm x})\cdot\left({\bm\phi}({\bm
x})\times
   {\bm\phi}^*({\bm x})\right)\quad.
\label{eq:A.5}
\eea
\end{widetext}
Here ${\bm m}$ and ${\bm\phi}$ are the magnetic and the superconducting order
parameter, respectively, and the latter has been represented as a complex
3-vector in spin space.\cite{Vollhardt_Woelfle_1990} $t_m$ and $t_{\Delta}$ are
the dimensionless distances from the magnetic and superconducting critical
point, and $\xi_m$ and $\xi_{\Delta}$ are the (zero-temperature) magnetic and
superconducting coherence lengths, respectively. ${\tilde g}$ is a
phenomenological coupling constant. At the mean-field level, ${\bm m} =
(0,0,{\tilde m})$, and ${\bm\phi} = ({\tilde\Delta},i{\tilde\Delta},0)$, with
${\tilde m}\propto m$ and ${\tilde\Delta}\propto\Delta_{\uparrow}$, and the
theory is seen to have the structure of the Landau theory derived from the
miscroscopic theory in Sec.\ \ref{app:A.1} above.

This action is invariant under a symmetry group SO(3)$\times$U(1), which
represents co-rotations of the vectors ${\bm m}$ and ${\bm\phi}$, and in
addition a gauge transformation of the complex field ${\bm\phi}$. In the
ordered state given above, this symmetry is spontaneously broken, and the
ordered state is invariant only under rotations about the 3-axis. We thus
expect dim(SO(3)$\times$U(1)/SO(2)) = 3 Goldstone modes. An explicit
calculation, involving an expansion to Gaussian order about the ordered state,
confirms this expectation. Of the three soft modes, two are spin wave-like
modes, namely, superpositions of the transverse components of ${\bm m}$ and the
3-component of ${\bm\phi}$, while the third is an Anderson-Bogoliubov-like mode
that reflects the broken gauge symmetry. Specifically, the transverse ${\bm
m}$-susceptibility has an overlap with one of the spin-wave Goldstone modes.
The Gaussian theory yields
\bea
{\tilde\chi}_{\text{T}}({\bm p}) &\equiv& \langle m_1({\bm p})\,m_1(-{\bm
p})\rangle
\nonumber\\
   &=& \frac{{\tilde m}\xi_m^{-2}\,\left({\tilde g}{\tilde m}\xi_{\Delta}^{-2}
           + {\bm p}^2\right)}
   {2{\bm p}^2\,\left[{\tilde g}\left({\tilde\Delta}^2\xi_m^{-2} +
      {\tilde m}^2\xi_{\Delta}^{-2}\right) + {\tilde m}{\bm p}^2\right]}
\label{eq:A.6}
\eea

From this result we see that, in the momentum range
\be
{\tilde g}\,{\tilde m}\,\xi_{\Delta}^{-2} < {\bm p}^2 < {\tilde
g}\,({\tilde\Delta}^2/{\tilde m})\,\xi_m^{-2}\quad,
\label{eq:A.7}
\ee
the transverse susceptibility is effectively massive,
\be
{\tilde\chi}_{\text{T}}({\bm p}) \approx {\tilde m}/2{\tilde
g}{\tilde\Delta}^2\quad.
\label{eq:A.8}
\ee

${\tilde m}$ and ${\tilde\Delta}$ can be related to $m$ and
$\Delta_{\uparrow}$, respectively, in the microscopic theory by comparing Eq.\
(\ref{eq:A.5}) with the Landau theory in Sec. \ref{app:A.1} above. One finds
the correspondence
\bse
\label{eqs:A.9}
\bea
{\tilde m} \cong m\,\sqrt{\Gamma_t/2T}\quad,
\label{eq:A.9a}\\
{\tilde\Delta} \cong \Delta_{\uparrow}\,\sqrt{N_{\text{F}}/T}\quad.
\label{eq:A.9b}
\eea
${\tilde g}$ does not corresponds to a simple constant in the microscopic
theory, but rather to a function of $m$ (or $\delta = \Gamma_{\text{t}}\,m$)
and the function $g(\delta/2T)$ in Eq.\ (\ref{eq:3.5c}),
\be
{\tilde g} \cong \sqrt{2T\Gamma_{\text{t}}}\ \frac{1}{3\delta}\ g(\delta/2T)
   \quad.
\label{eq:A.9c}
\ee
\ese
This complicated behavior of the effective coupling between the magnetic and
superconducting order parameters is not reflected by the LGW theory. It is a
result of the itinerant nature of the electrons, and the corresponding soft
particle-hole excitations, as is discussed in Sec.\ \ref{sec:V}. Upon
substituting Eqs.\ (\ref{eqs:A.9}) in Eq.\ (\ref{eq:A.8}) we see that the
latter is the same result as Eq.\ (\ref{eq:3.5a}) for the magnetic
susceptibility.

The momentum range given by Eq.\ (\ref{eq:A.7}) exists only if the condition
$\Delta\gg{\tilde m}\xi_m/\xi_{\Delta}$ is fulfilled. In terms of the Stoner
gap $\delta = \Gamma_{\text{t}}\,m$ and $\Delta_{\uparrow}$ this condition
takes the form
\be
\delta/\Delta_{\uparrow} \ll \xi_{\Delta}/\xi_{m}\quad.
\label{eq:A.10}
\ee
These results are further discussed in Sec.\ \ref{sec:V}.

\section{The spin susceptibility}
\label{app:B}

In this appendix we establish the relation between the physical spin
susceptibility $\chi_{\text{s}}$, as measured, e.g., by neutron scattering, and
the quantity $\chi$ that emerges as the $\delta\bm{M}$ propagator of our field
theory, see Eqs.\ (\ref{eq:2.9}, \ref{eqs:2.10}). We start by adding a magnetic
field $\bm{h}(x)$ to our action that couples linearly to the electron spin
density and acts as a source field for spin-density correlation functions. The
action, Eq.\ (\ref{eq:2.1}), becomes \bea {\cal A}[{\bm M},{\cal
G},\Lambda;\bm{h}] &=&
    \Tr(\Lambda{\cal G}) - \int dx\ {\bm M}(x)\cdot {\bm M}(x)
\nonumber\\
&&\hskip -75pt + \frac{1}{2}\Tr
   \ln ({\tilde G}_{0}^{-1} + \sqrt{2\Gamma_{\text t}}
   {\bm\gamma}\cdot{\bm M} - \Lambda^{\text T} + \bm{\gamma}\cdot\bm{h})\quad.
\label{eq:B.1}
\eea
The partition function,
\be
Z[\bm{h}] = \int D[\delta\bm{M},\delta\lambda,\delta\mathcal{G}]\
            e^{{\cal A}[{\bm M},{\cal G},\Lambda;\bm{h}]}\quad,
\label{eq:B.2}
\ee
then serves as the generating functional for the magnetization,
\be
m = \langle n_{\text{s}}^3(x)\rangle = \frac{\delta}{\delta h_3(x)}
    \Biggl\vert_{\bm{h}=0}\ \ln Z[h]\quad,
\label{eq:B.3}
\ee
and for the spin susceptibility,
\be
\chi_{\text{s}}^{ij}(x-y) = \langle \delta n_{\text{s}}^i(x)\,
                                       \delta n_{\text{s}}^j(y)\rangle
   = \frac{\delta^2}{\delta h_i(x)\,\delta h_j(y)}\Biggl\vert_{\bm{h}=0}\
                \hskip -5pt \ln Z[h]\ ,
\label{eq:B.4}
\ee
where $\bm{n}_{\text{s}}(x)$ is the electron spin density, and
$\delta n_{\text{s}}^i(x) = n_{\text{s}}^i(x) - \langle n_{\text{s}}^i(x)
\rangle$.

The simplest way to deal with the source field is to shift $\bm{M}$:
$\bm{M}(x) \rightarrow \bm{M}(x) - \bm{h}(x)/\sqrt{2\Gamma_{\text{t}}}$.
We then find
\be
m = \sqrt{2/\Gamma_{\text{t}}}\,\langle M_3(x)\rangle\quad,
\label{eq:B.5}
\ee
in agreement with Eq.\ (\ref{eq:2.5a}). An evaluation of
$\langle M_3(x)\rangle$ to one-loop order yields the magnetic equation
of state, Eq.\ (\ref{eq:2.14}). Similarly,
\be
\chi_{\text{s}}^{ij}(x-y) = \langle\delta M_i(x)\,\delta M_j(y)\rangle
   - \delta_{ij}\,\delta(x-y)\quad.
\label{eq:B.6}
\ee
If we evaluate the $\delta \bm{M}$-correlation function in Gaussian
approximation, we find\cite{shift_footnote}
\be \chi_{\text{s}}^{ij}(x-y) \approx \left[
   \chi_{ij}(x-y) - \delta_{ij}\,\delta(x-y)\right]/\Gamma_{\text{t}}\quad.
\label{eq:B.7}
\ee
We see that the $\delta \bm{M}$-propagator of the field theory is simply
related to the physical spin susceptibility. In particular, the transverse part
of $\chi_{\text{s}}$ is massive if and only if that of $\chi$ is. For the
Fourier transform of the longitudinal part, we find from Eq.\ (\ref{eq:B.7})
\be \chi_{\text{s}}^{\text{L}}(k) \equiv \chi_{\text{s}}^{33}(k)
   = \frac{-\chi_{\text{L}}^0(k)}{1 + \Gamma_{\text{t}}\chi_{\text{L}}^0(k)}
      \quad,
\label{eq:B.8}
\ee
which has the expected RPA-type structure. Notice that the ``contact term'',
which results from the term in $\mathcal{A}$ that is quadratic in $\bm{h}$,
provides the numerator in Eq.\ (\ref{eq:B.8}) that is missing in
$\chi_{\text{L}}$.


\begin{thebibliography}{42}
\bibitem[{flu()}]{flux_lattice_footnote}
\bibinfo{note}{In addition, the superconducting state cannot be uniform, but
  must display a spontaneous flux-lattice structure, with both the magnetic and
  the superconducting order parameters showing spatial modulation. We will
  ignore this effect in this paper and come back to it in a separate
  publication.}
\bibitem[{err()}]{errors_footnote}
\bibinfo{note}{While the end result in I was correct, the derivation contained
  a number of compensating errors.}
\bibitem[{dia()}]{diagram_rule_footnote}
\bibinfo{note}{This is closely related to the diagram rules in Sec. IV.A of
  Ref.\ \protect\onlinecite{us_fermions_I}.}
\bibitem[{mis()}]{missing_factor_footnote}
\bibinfo{note}{Equation (3.10) in I had a factor of $T/2V$ missing.}
\bibitem[{met()}]{method_footnote}
\bibinfo{note}{It is not obvious what the free energy should be minimized with
  respect to, and what should be kept constant in such a procedure. While the
  end result in I was correct, the derivation contained a number of
  compensating errors. This is the main reason why Ma's method is advantageous
  compared to a minimization of the free energy.}
\bibitem[{1st()}]{1st_order_footnote}
\bibinfo{note}{This assumes that the transition is of second order. Close to
  the ferromagnetic quantum critical point, this may not be the case, see Ref.
  \onlinecite{Chubukov_et_al_2003}}.
\bibitem[{num()}]{numerical_mistake_footnote}
\bibinfo{note}{The code used to calculate $T_{\text{c}\uparrow}$ in zero-loop
  approximation in I had the wave number in $\chi_{\text{L}}$ pinned to the
  paramagnetic Fermi surface rather than to the up-spin Fermi surface. As a
  consequence, our present results for $T_{\text{c}\uparrow}$ differ slightly
  from those in I for the same parameter values.}
\bibitem[{del()}]{delta_footnote}
\bibinfo{note}{In the elemental ferromagnets Fe and Co the exchange splitting
  $\delta$ is larger than the Curie temperature by a factor on the order of
  10.\protect\cite{Kuebler_2000} For the superconducting ferromagnets of
  interest here we are not aware of any values for $\delta$.}
\bibitem[{qua()}]{quartic_term_footnote}
\bibinfo{note}{In addition to the quartic term $\vert{\bm\phi}\vert^4$, there
  is in general also a term $({\bm\phi}\times{\bm\phi}^*)^2$. Adding this term
  does not change our conclusions.}
\bibitem[{shi()}]{shift_footnote}
\bibinfo{note}{Note that this relation holds both before and after the shifts
  that diagonalize the Gaussian action. The reason is the same cancellation
  that leads to the diagram rule in Sec.\ \protect\ref{subsec:II.C}.}
\end{thebibliography}

\end{document}